\documentclass[conference]{IEEEtran}
\IEEEoverridecommandlockouts
\usepackage{scrextend}
\usepackage[utf8]{inputenc}
\usepackage{bbm}
\usepackage{bm}
\usepackage{xcolor}
\usepackage{multirow}
\usepackage{bbold}
\usepackage{siunitx}
\usepackage{mathtools}
\usepackage{enumitem}
\usepackage{algpseudocode}
\usepackage{booktabs}
\usepackage{xfrac}
\usepackage[linesnumbered,ruled,vlined]{algorithm2e}
\allowdisplaybreaks

\usepackage{mathtools}

\DeclarePairedDelimiter\floor{\lfloor}{\rfloor}

\usepackage{cite}
\makeatletter
\def\ps@IEEEtitlepagestyle{%
\def\@evenfoot{}%
}

\let\Right\right
\let\Left\left
\makeatletter
\def\right#1{\Right#1\@ifnextchar){\!\right}{}}
\def\left#1{\Left#1\@ifnextchar({\!\left}{}}
\makeatother

\DeclarePairedDelimiter\abs{\lvert}{\rvert}

\ifCLASSOPTIONcompsoc
    \usepackage[caption=false, font=normalsize, labelfont=sf, textfont=sf]{subfig}
\else
\usepackage[caption=false, font=footnotesize]{subfig}
\fi

   \usepackage{graphicx}

\usepackage{multirow}

\begin{document}
\title{Conservative Bias Linear Power Flow Approximations: Application to Unit Commitment
\thanks{Paprapee Buason (email: paprapee.b@rmutp.ac.th), 
Sidhant Misra (email: sidhant@lanl.gov), and 
Daniel K. Molzahn (email: molzahn@gatech.edu).

This work is supported by the Advanced Grid Modeling (AGM) Program of the U.S. Department of Energy, Office of Electricity, and NSF Award \#2145564.}
}

\author{
\IEEEauthorblockN{Paprapee Buason\IEEEauthorrefmark{1}\IEEEauthorrefmark{2},
Sidhant Misra\IEEEauthorrefmark{1},
Daniel K. Molzahn\IEEEauthorrefmark{3}}

\IEEEauthorblockA{\IEEEauthorrefmark{1}
\textit{Los Alamos National Laboratory}, Los Alamos, NM, USA}

\IEEEauthorblockA{\IEEEauthorrefmark{2}
\textit{Department of Mechatronics Engineering}, 
\textit{Rajamangala University of Technology Phra Nakhon}, Bangkok, Thailand}

\IEEEauthorblockA{\IEEEauthorrefmark{3}
\textit{School of Electrical and Computer Engineering}, 
\textit{Georgia Institute of Technology}, Atlanta, GA, USA}
}

\maketitle

\begin{abstract}
Accurate modeling of power flow behavior is essential for a wide range of power system applications, yet the nonlinear and nonconvex structure of the underlying equations often limits their direct use in large-scale optimization problems. As a result, linear models are frequently adopted to improve computational tractability, though these simplifications can introduce excessive approximation error or lead to constraint violations. This paper presents a linear approximation framework, referred to as \emph{Conservative Bias Linear Approximations (CBLA)}, that systematically incorporates conservativeness into the approximation process. Rather than solely minimizing local linearization error, CBLA constructs linear constraints that bound the nonlinear functions of interest over a defined operating region while reducing overall approximation bias. The proposed approach maintains the simplicity of linear formulations and allows the approximation to be shaped through user-defined loss functions tailored to specific system quantities. Numerical studies demonstrate that CBLA provides more reliable and accurate approximations than conventional linearization techniques, and its integration into a unit commitment formulation results in improved feasibility and reduced operating costs.

\end{abstract}


\section{Introduction} \label{sec:introduction}

Accurate representation of network physics is fundamental to a broad range of power system optimization problems. The nonlinear alternating current (AC) power flow equations govern voltage magnitudes, phase angles, and line flows, and therefore directly influence planning, operational scheduling, and reliability assessments. These equations appear in applications including resilient infrastructure design~\cite{gupta_buason_molzahn-fairness_pv_limit, owen_aquino_talkington_molzahn-EVacuation_feeder, bhusal2020}, AC unit commitment (UC)~\cite{castillo2016, Sampath2018}, and bilevel or market-based formulations~\cite{buason2023datadriven, wogrin2020}. However, embedding the full nonlinear equations into optimization models introduces nonconvexities that significantly complicate computation, particularly in large-scale or mixed-integer settings.

To mitigate these computational challenges, numerous linear approximations of the power flow equations have been developed. Widely used models such as DC power flow~\cite{stott2009}, LinDistFlow~\cite{baran1989}, and first-order Taylor expansions~\cite{wood2013power} provide tractable representations that enable efficient optimization. While these approximations are attractive due to their simplicity, they are typically derived under restrictive assumptions, for example, small voltage angle differences or near-nominal voltage magnitudes. When system operating conditions deviate from these assumptions, the resulting linear models may produce substantial errors, potentially leading to suboptimal or infeasible solutions when evaluated against the underlying AC equations. Thus, a fundamental trade-off arises between computational tractability and physical accuracy.

Recognizing these limitations, recent research has investigated adaptive linearization strategies that tailor approximations to a specific system and operating region. Rather than relying on globally valid but coarse assumptions, these methods construct linear models using information obtained from representative operating samples. Optimization-based techniques~\cite{misra2018optimal} and sample-driven approaches~\cite{BUASON2022, buason2024adaptive, CHEN2022108573} compute linear coefficients by minimizing deviations from AC power flow solutions within a prescribed range of power injections; see~\cite{jia2023tutorial1, 10202779, jia2023tutorial2} for surveys of this emerging direction. By shifting computational effort to an offline stage where samples are generated and processed, adaptive approximations can provide improved fidelity while preserving linear structure for subsequent optimization tasks.

Among these approaches, the conservative linear approximation (CLA) framework~\cite{BUASON2022, buason2024adaptive} incorporates an additional structural property: directional conservativeness. In CLA, the linear approximation is constructed so that, across the sampled operating region, quantities of interest are consistently overestimated or underestimated relative to the AC solution. This feature is particularly valuable in optimization problems where certain types of approximation error are more harmful than others. For example, underestimating line currents in a thermal limit constraint may falsely indicate feasibility, whereas overestimation results primarily in conservativeness. By enforcing one-sided error behavior, CLA reduces the risk of infeasible solutions. However, strict conservativeness may increase overall approximation error, especially in regions where the nonlinear relationship is difficult to represent accurately with a single linear model.

Motivated by this observation, this paper introduces a generalized framework termed \emph{conservative bias linear approximation} (CBLA). The CBLA methodology retains the sample-based construction process but relaxes strict one-sided enforcement of conservativeness. Instead of imposing conservativeness as a hard constraint, CBLA incorporates it through a bias-aware error function that penalizes violations asymmetrically. This formulation enables a controlled balance between directional bias and approximation accuracy. As a result, CBLA can reduce excessive conservativeness while still discouraging the most critical forms of approximation error.

A distinguishing feature of CBLA is its flexibility in error function design. Users may define loss functions that reflect the operational priorities of a given application, allowing the approximation to emphasize particular quantities or error directions. This flexibility makes the framework adaptable to diverse optimization contexts, including those in which limited constraint violations are tolerable, such as chance-constrained or risk-aware formulations. By tailoring the regression objective to the needs of the downstream problem, CBLA produces linear models that are specialized to both the system and the intended optimization task.

This work extends our preliminary study in~\cite{buason_cbla}, where the CBLA concept was first introduced for power flow approximation. In the present journal version, we expand the framework and demonstrate its practical implications through extensive numerical evaluation. In particular, we integrate CBLA into a UC formulation~\cite{Padhy2004}, compare its performance with alternative linearization techniques, and describe a structured solution procedure for solving the resulting problem. The results illustrate that CBLA achieves improved feasibility and reduced operating cost relative to conventional linear models.

The primary contributions of this paper are summarized as follows:
\begin{enumerate}[label=(\textit{\roman*})]
\item A CBLA framework that balances directional conservativeness and approximation accuracy over a specified operating range.
\item A discussion of error function selection and its role in shaping the trade-off between bias and fidelity.
\item Numerical validation of CBLA across multiple test systems.
\item An application of CBLA to UC, including comparative results against established linear power flow models.
\end{enumerate}

The remainder of this paper is organized as follows. Section~\ref{sec:background} reviews relevant background on power flow modeling and conservative linear approximations. Section~\ref{sec:CBLA} presents the proposed CBLA methodology. Section~\ref{sec:uc} details the UC formulation and solution approach. Section~\ref{sec:simulation} provides numerical results, and Section~\ref{sec:future work} concludes the paper.


\section{Background on AC Power Flow and Conservative Linear Approximations} \label{sec:background}

This section reviews the AC power flow model and summarizes the sample-based conservative linear approximation (CLA) framework that motivates the development of the proposed approach.

\subsection{AC Power Flow Model} \label{sub:PF_equations}

We consider a power network in which one bus is selected as the reference with voltage angle fixed at zero. Let $V_i$ and $\theta_i$ denote the voltage magnitude and phase angle at bus $i$, respectively. The active and reactive power injections are represented by $P_i$ and $Q_i$. Subscripts $(\cdot)_i$ refer to quantities associated with bus $i$, while $(\cdot)_{ik}$ indicates quantities related to the branch connecting buses $i$ and $k$.

The steady-state AC power flow equations at bus $i$ are given by
\begin{subequations}
\label{eq:power_flow}
\begin{align}
	P_i &= V_i^2 G_{ii} + \sum_{k \in \mathcal B_i} V_i V_k \left(G_{ik}\cos \theta_{ik} + B_{ik}\sin \theta_{ik}\right), \label{eq:dP} \\
	Q_i &= -V_i^2 B_{ii} + \sum_{k \in \mathcal B_i} V_i V_k \left(G_{ik}\sin \theta_{ik} - B_{ik}\cos \theta_{ik}\right), \label{eq:dQ}
\end{align}
\end{subequations}
where $\theta_{ik} := \theta_i - \theta_k$, and $G$ and $B$ denote the real and imaginary parts of the network admittance matrix. The set $\mathcal B_i$ contains bus $i$ and all buses directly connected to it.

The nonlinear trigonometric and multiplicative terms in~\eqref{eq:power_flow} introduce nonconvexities that significantly complicate optimization problems embedding AC network constraints. This motivates the construction of tractable linear representations of selected quantities derived from these equations.

\subsection{Conservative Linear Approximations} \label{sub:cla}

To improve tractability while preserving feasibility with respect to the AC model, we previously developed a sample-based CLA framework~\cite{BUASON2022}. The goal of CLA is to construct affine functions of power injections that approximate nonlinear quantities of interest—such as voltage magnitudes or current flows—while ensuring one-sided error behavior over a prescribed operating region.

The CLA procedure begins by defining an operating set $\mathcal S$ for power injections. For example, load demands may be sampled from a probability distribution $\mathbb{P}_{\mathcal S}$ over
\[
\mathcal{S} = \{ P_{L_d}^{\min} \le P_{L_d} \le P_{L_d}^{\max}, \;
Q_{L_d}^{\min} \le Q_{L_d} \le Q_{L_d}^{\max},
\; \forall L_d \in \mathcal N_D \},
\]
where $\mathcal N_D$ denotes buses with loads and the superscripts indicate lower and upper bounds. A uniform distribution is commonly adopted. For each sampled injection vector, the AC power flow equations are solved to obtain the corresponding nonlinear quantities of interest. These samples are then used to compute an affine approximation through constrained regression.

\begin{figure}[ht!]
\vspace{-1em}
	\centering 
	\includegraphics[trim=0.5cm 0.5cm 0.2cm 0.8cm, clip, width=1.05\linewidth]{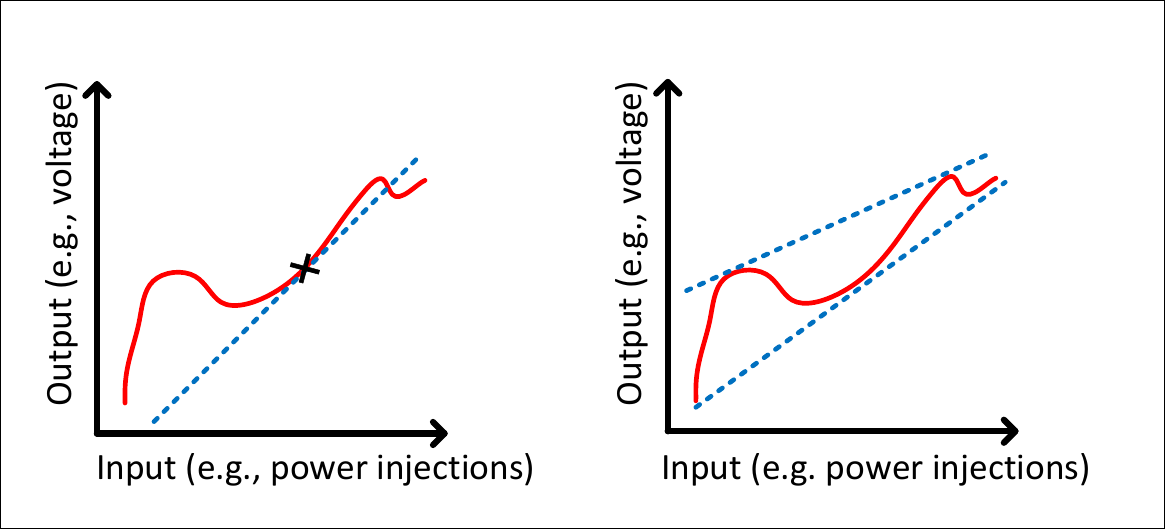} 
	\caption{Comparison of a conventional linearization (left) and CLAs (right). The solid curve represents the nonlinear function. The dotted line on the left corresponds to a first-order Taylor approximation at the marked point. The dotted lines on the right represent over- and under-estimating conservative approximations.}
	\label{fig:cla_fig}
    \vspace{-1em}
\end{figure}

An illustrative comparison between a conventional first-order Taylor expansion and a CLA is shown in Fig.~\ref{fig:cla_fig}. Unlike a local linearization centered at a single operating point, a CLA is constructed to systematically overestimate or underestimate the nonlinear function throughout the sampled region.

In addition to improved regional fidelity, CLAs provide two practical advantages. First, each quantity of interest can be approximated independently, enabling parallel computation of coefficients. Second, if conservativeness is maintained, nonlinear inequality constraints can be replaced with linear inequalities while preserving feasibility relative to the AC model. This enables compatibility with commercial linear and mixed-integer solvers.

Let $\gamma$ denote a scalar quantity derived from the AC power flow solution, such as a voltage magnitude or branch current magnitude. Using vector notation for power injections, an overestimating CLA takes the affine form
\begin{equation}
\label{eq:cla_template}
a_0 + \bm{a}_1^T
\begin{bmatrix}
\bm{P} \\
\bm{Q}
\end{bmatrix},
\end{equation}
where $a_0$ and $\bm{a}_1$ are regression coefficients. Conservativeness requires that, for injections within the operating range,
\begin{equation}
\gamma \le
a_0 + \bm{a}_1^T
\begin{bmatrix}
\bm{P} \\
\bm{Q}
\end{bmatrix}.
\label{eq:cla_setup}
\end{equation}

If~\eqref{eq:cla_setup} holds, then an upper bound constraint $\gamma \le \gamma^{\max}$ can be enforced through the linear condition
\[
a_0 + \bm{a}_1^T
\begin{bmatrix}
\bm{P} \\
\bm{Q}
\end{bmatrix}
\le \gamma^{\max},
\]
without explicitly embedding the nonlinear AC equations. Thus, feasibility with respect to the nonlinear model is preserved while maintaining linear structure.

The coefficients in~\eqref{eq:cla_template} are obtained by solving the constrained regression problem
\begin{subequations}
\label{eq:regression}
\begin{align}
\min_{a_0,\, \bm{a}_1} \quad &
\frac{1}{M}
\sum_{m=1}^{M}
\mathcal{L}
\left(
\gamma_m -
\left(
a_0 + \bm{a}_1^T
\begin{bmatrix}
\bm{P}_m \\
\bm{Q}_m
\end{bmatrix}
\right)
\right) \label{eq:regression_unconstrained} \\
\text{s.t.} \quad &
\gamma_m -
\left(
a_0 + \bm{a}_1^T
\begin{bmatrix}
\bm{P}_m \\
\bm{Q}_m
\end{bmatrix}
\right)
\le 0,
\quad m=1,\dots,M. \label{eq:regression_over}
\end{align}
\end{subequations}
where $m$ indexes samples, $M$ denotes the number of samples, and $\mathcal{L}(\cdot)$ is a loss function such as the absolute value ($\ell_1$) or squared error ($\ell_2$). Underestimating CLAs are obtained by reversing the inequality in~\eqref{eq:regression_over}.

While the constraint in~\eqref{eq:regression_over} guarantees directional conservativeness, it restricts the feasible regression set and may increase approximation error compared to the unconstrained formulation obtained by removing this condition. Consequently, strict conservativeness can reduce overall accuracy, particularly in regions where the nonlinear relationship is difficult to capture with a single affine model.

The next section introduces the main contribution of this paper, which extends the CLA framework by incorporating directional bias directly into the regression objective. This formulation enables a more flexible trade-off between feasibility protection and approximation accuracy.

\section{Conservative Bias Linear Approximations} \label{sec:CBLA}

The CLA framework described in Section~\ref{sub:cla} guarantees strict one-sided conservativeness across all sampled operating points. While this property is desirable for feasibility protection, enforcing it uniformly may lead to unnecessarily large approximation errors when certain samples are difficult to represent accurately with a single affine model. 

To address this limitation, we introduce a sample-based \emph{conservative bias linear approximation} (CBLA) methodology. Similar to CLA, the proposed approach is adaptive in that it is constructed using samples drawn from a specified operating region for a particular system. However, instead of imposing conservativeness as a hard constraint, CBLA incorporates a bias toward conservativeness directly into the optimization objective. This formulation enables the approximation to remain predominantly conservative while allowing limited violations at a controlled cost. As a result, CBLA seeks to achieve a principled balance between directional bias and approximation accuracy.

\subsection{Formulation} \label{sub:cbla_formulation}

Let $\epsilon_m$ denote the approximation error (mismatch) at sample $m$, defined as the difference between the true quantity of interest and its affine approximation. The CBLA coefficients are computed by solving the optimization problem
\begin{equation}
\label{eq:cbla}
\min_{a_0,\, \bm{a}_1} \quad \frac{1}{M} \sum_{m=1}^{M} f(\epsilon_m),
\end{equation}
where
\begin{align}
[\forall m = 1,\ldots, M] \notag \\
\epsilon_m 
= \gamma_m 
- \left(
a_0 + \bm{a}_1^T
\begin{bmatrix}
\bm{P}_m \\
\bm{Q}_m
\end{bmatrix}
\right).
\label{eq:mismatch}
\end{align}

The function $f(\epsilon_m)$ determines how errors are penalized and is defined piecewise as
\begin{align}
f(\epsilon_m) =
\begin{dcases}
g(\epsilon_m), & \text{if } \epsilon_m \le 0, \\
h(\epsilon_m), & \text{otherwise.}
\end{dcases}
\label{eq:error_function}
\end{align}

The optimization problem in~\eqref{eq:cbla} therefore minimizes the average penalized error across all samples by selecting coefficients $a_0$ and $\bm{a}_1$. The key distinction from CLA lies in the structure of the error function. Rather than forbidding violations of conservativeness, CBLA assigns asymmetric penalties based on the sign of $\epsilon_m$.

For an overestimating approximation, violations correspond to $\epsilon_m > 0$. In this case, $h(\epsilon_m)$ is chosen to impose a relatively large penalty, while $g(\epsilon_m)$ remains comparatively small for non-violating samples. For underestimating approximations, the roles of $g(\cdot)$ and $h(\cdot)$ are reversed. This asymmetric design induces a \emph{tendency} toward conservativeness without strictly enforcing it.

\subsection{Error Function Design} \label{sub:cbla_error_function}

The selection of the error function $f(\cdot)$ plays a central role in determining the trade-off between conservativeness and approximation accuracy. Its design may depend on system requirements, the specific quantity being approximated, and the acceptable level of directional violation.

To illustrate the relationship between CLA and CBLA, consider the CLA regression problem in~\eqref{eq:regression} with an $\ell_1$ loss. That formulation can be interpreted within the CBLA framework by defining
\begin{align}
f(\epsilon_m) =
\begin{dcases}
\epsilon_m, & \text{if } \epsilon_m \le 0, \\
\infty, & \text{otherwise.}
\end{dcases}
\label{eq:error_function_cla}
\end{align}

\begin{figure}[ht!]
\vspace{-0.8em}
	\centering 
	\includegraphics[trim=0.5cm 0.1cm 0.4cm 0.3cm, clip, width=0.57\linewidth]{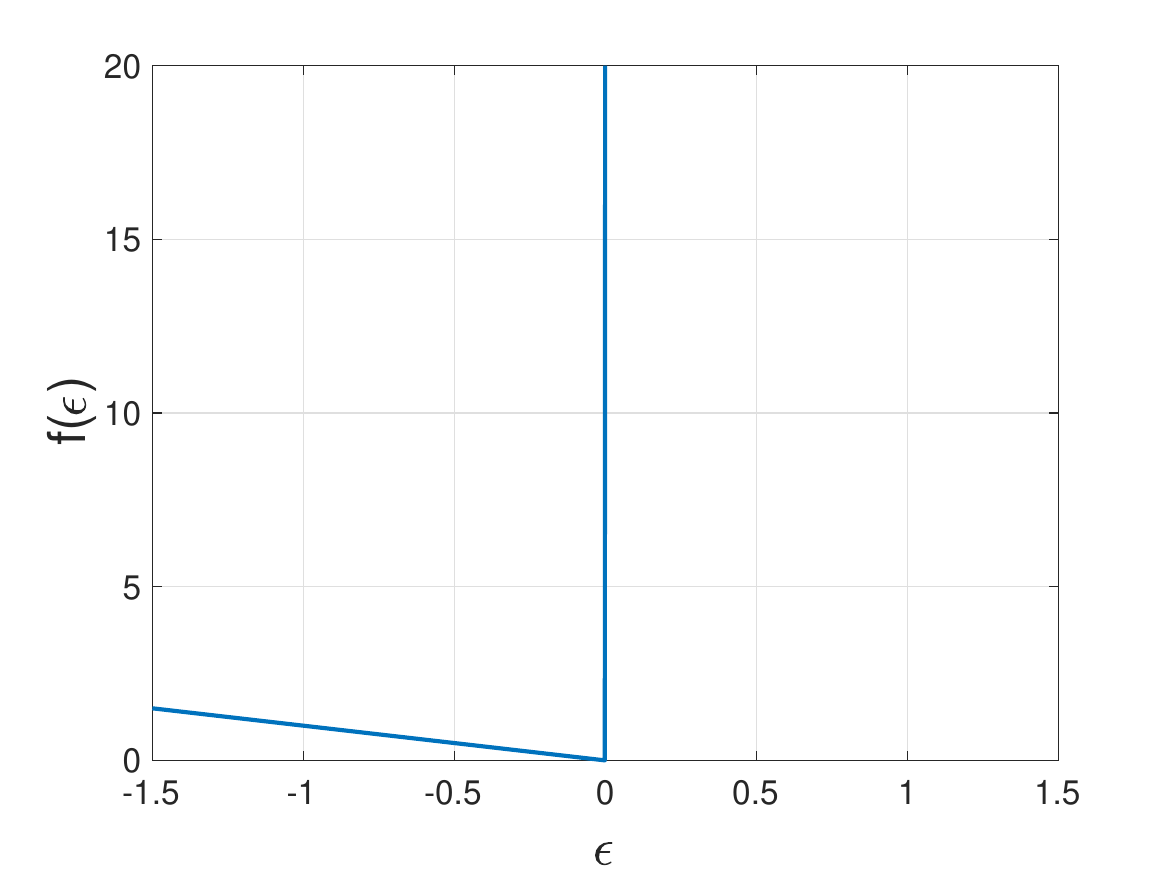} 
	\caption{Error function corresponding to CLA, where $g(\epsilon)=\epsilon$ and $h(\epsilon)=\infty$.}
	\label{fig:error_function_cbla}
 \vspace{-0.5em}
\end{figure}

As shown in Fig.~\ref{fig:error_function_cbla}, assigning infinite cost to $\epsilon_m>0$ strictly prohibits violations, thereby recovering the CLA formulation. In contrast, CBLA replaces the infinite penalty with a finite but potentially large function, permitting limited violations when doing so improves overall accuracy.

\begin{figure}[ht!]
\vspace{-0.8em}
	\centering 
	\includegraphics[trim=0.5cm 0.1cm 0.4cm 0.3cm, clip, width=0.57\linewidth]{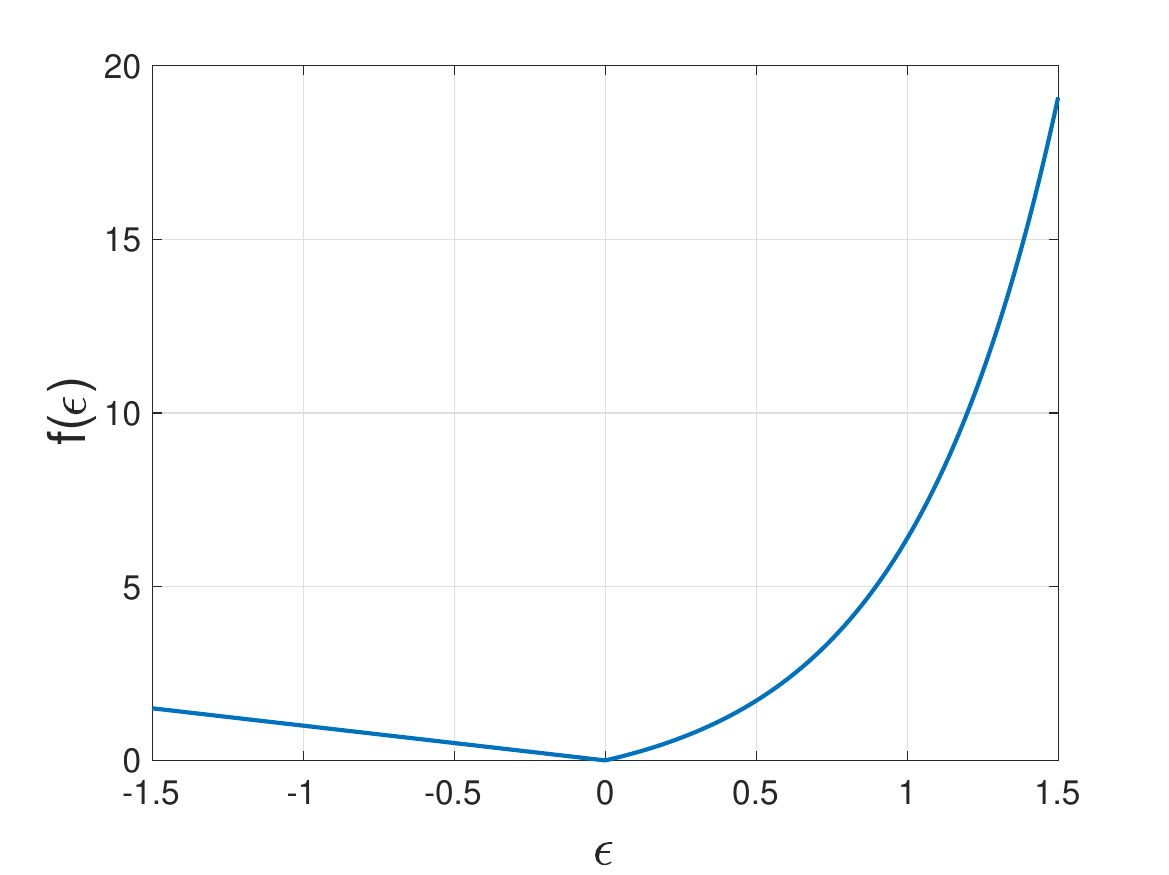} 
	\caption{Example CBLA error function for an overestimating approximation, where $g(\epsilon)=\epsilon$ and $h(\epsilon)=e^{2\epsilon}-1$.}
	\label{fig:error_function_example}
 \vspace{-0.5em}
\end{figure}

An example is shown in Fig.~\ref{fig:error_function_example}, where $g(\epsilon)=\epsilon$ and $h(\epsilon)=e^{2\epsilon}-1$. In this case, positive errors incur exponentially increasing penalties, discouraging large violations while allowing small deviations when necessary. Increasing the growth rate of $h(\epsilon)$ promotes stronger conservativeness, whereas reducing its slope yields a more accurate but less conservative approximation.

When both $g(\epsilon)$ and $h(\epsilon)$ are linear functions, i.e., the error function is piecewise linear, the CBLA regression problem can be reformulated as a linear program:
\begin{subequations}
\label{eq:regression_linear}
\begin{align}
\min_{a_0,\, \bm{a}_1} \quad &
\frac{1}{M}\sum_{m=1}^{M} z_m \\
\text{s.t.} \quad 
&[\forall m = 1,\ldots,M] \notag \\
&\epsilon_m = \gamma_m 
- \left(
a_0 + \bm{a}_1^T
\begin{bmatrix}
\bm{P}_m \\
\bm{Q}_m
\end{bmatrix}
\right), \label{eq:mismatch_linear} \\
&z_m \ge k_1 \epsilon_m, \label{eq:linear_constraint1} \\
&z_m \ge k_2 \epsilon_m, \label{eq:linear_constraint2}
\end{align}
\end{subequations}
where $z_m$ is an auxiliary variable and $k_1$, $k_2$ define the slopes of $g(\epsilon)=k_1\epsilon$ and $h(\epsilon)=k_2\epsilon$, respectively.

If nonlinear penalty functions are selected, the resulting optimization problem becomes a nonlinear program (or potentially a mixed-integer nonlinear program, depending on the formulation). Such problems can be implemented using user-defined functions in Julia and solved using packages such as \texttt{Optim}, which provides a flexible framework for constrained optimization~\cite{optimjl}.

\section{Unit Commitment} \label{sec:uc}
The unit commitment (UC) problem is an optimization problem in power system operations that aims to determine the optimal schedule of generating units over a specified time horizon, typically 24 hours. The primary goal is to minimize the total cost of operation while ensuring that the demand for electricity is met and the technical and operational constraints of the generation units and transmission network are satisfied. 

The variability and uncertainty associated with loads, particularly when load shedding may be necessary, and generation introduce additional complexities that the UC problem must address to ensure grid reliability and operational efficiency. As a mixed-integer optimization problem, the UC formulation involves binary variables to represent the on/off status of generators, along with continuous variables to model power output.

\subsection{Computational Procedure and Parallelization Potential} \label{sub:computation}

\begin{figure}[bh!]
\vspace{-0.8em}
	\centering
	\includegraphics[trim=0.4cm 0.3cm 0.4cm 0.3cm, clip, width=0.7\linewidth]{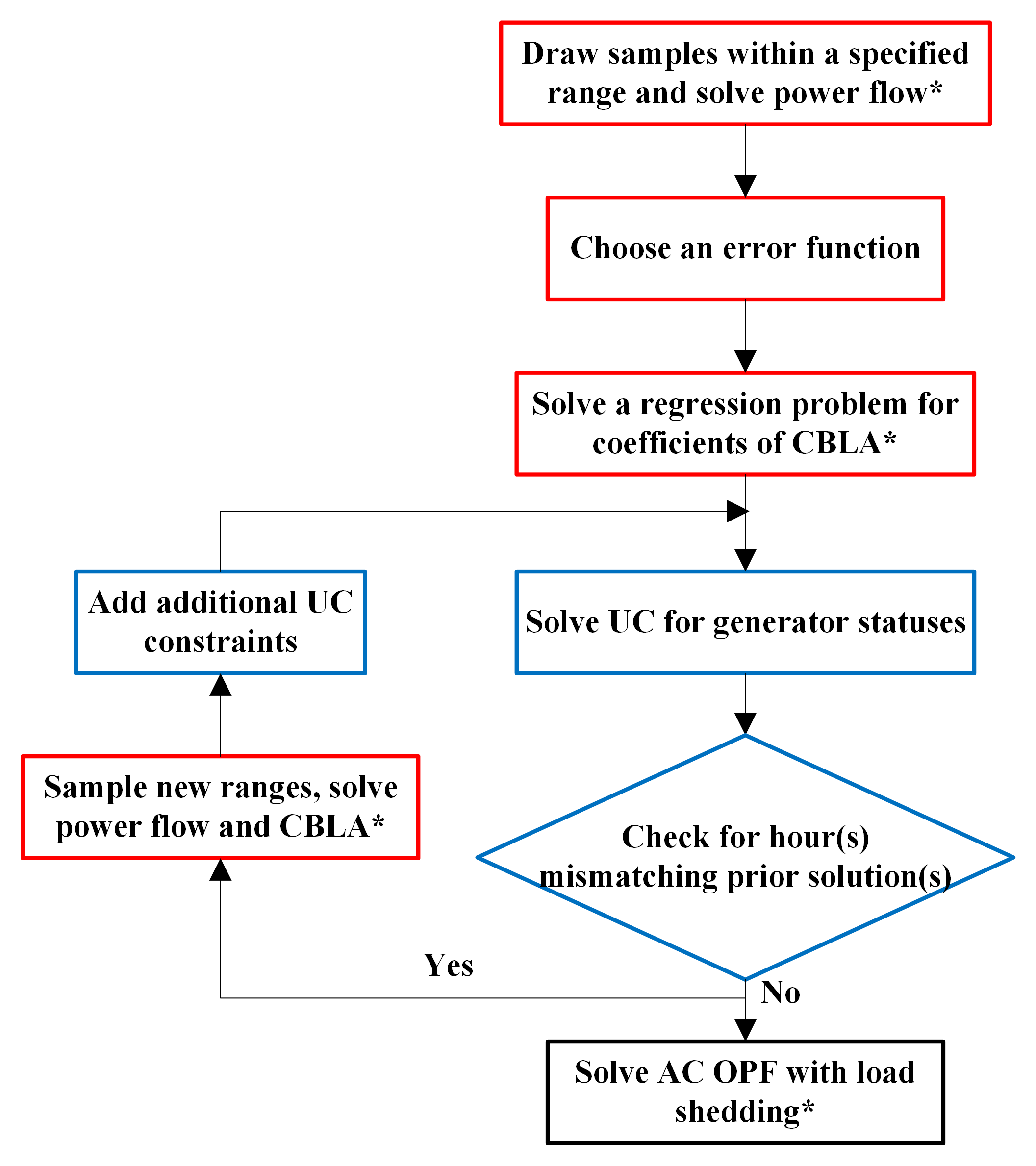}
	\caption{A flowchart illustrating the computational process for solving the UC problem using CBLA. Steps enclosed in red boxes correspond to CBLA, while steps enclosed in blue boxes relate to the UC problem. Steps marked with $*$ indicate parallelizable processes.}
	\label{fig:algorithm}
\end{figure}

Solving the UC problem using CBLA involves multiple computational steps, each contributing to accurately determining generator commitments and dispatch while leveraging efficient approximations of power flow equations. Given the complexity of UC and the nonlinearity of AC power flow, CBLA serves as an effective method to construct computationally efficient linear approximations. The overall process consists of three key steps:

\begin{enumerate}
    \item \textbf{CBLA computation}: Solve the CBLA formulation in~\eqref{eq:cbla} to approximate voltage and current over a range of operating conditions. This step requires generating samples within expected operating ranges and solving the CBLA optimization problem. \label{step:cbla_computation}
    \item \textbf{UC optimization}: Using the CBLA approximations obtained in Step~\ref{step:cbla_computation}, solve the UC problem to determine generator statuses for each generator at each hour while ensuring operational constraints are satisfied.\label{step:unit_commitment}
    \item \textbf{AC optimal power flow (AC OPF) evaluation}: Given the generator statuses from Step~\ref{step:unit_commitment}, solve the AC OPF that allows load shedding to determine the generation dispatch at each hour, ensuring feasibility under AC power flow constraints~\cite{Happ1977}. \label{step:ac_opf}
\end{enumerate}

This computational framework allows parallel execution, particularly in Steps~\ref{step:cbla_computation} and~\ref{step:ac_opf}. In Step~\ref{step:cbla_computation}, solving CBLA involves two key computations that can be parallelized. First, the power flow solutions at each operating point can be computed independently. Second, the regression problem for the CBLA can be solved independently for each quantity, using the same set of samples across all regression problems. Similarly, in Step~\ref{step:ac_opf}, since ramping constraints are not enforced and only minimum up- and down-time constraints are considered, the AC OPF computations at different time steps are independent once generator statuses are fixed, enabling parallel execution across time periods. Fig.~\ref{fig:algorithm} illustrates the full set of computational steps required to solve the UC problem using the CBLA approach.

\subsection{Optimization Problem Formulation}

Let $u_{g,t}$ be a generator on/off status and $s_{g,t} \; (r_{g,t})$ be a generator startup (shutdown) status for generator $g$ at time $t$.  The UC problem with the DC power flow~\cite{vandenbergh2014} is formulated as follows: 
\begin{subequations}
\label{eq:unit_commitment_main}
\begin{align} 
    \min \ & \sum_{t \in \mathcal{T}} \left( \sum_{g \in \mathcal{G}} c_g(P_{g,t})  +  C_{g}^\text{startup} s_{g,t} + C_{g}^\text{shutdown} r_{g,t} \right) \label{eq:objective} \\
    \text{s.t.} \ 
    & \sum_{g \in \mathcal{G}} P_{g,t} = D_t, \quad \forall t \in \mathcal{T}, & \label{eq:power_balance} \\
    & [\forall g \in \mathcal{G}] \notag \\
    & P_{g,t} \leq P_{g}^\text{max} u_{g,t}, \; P_{g,t} \geq P_{g}^\text{min} u_{g,t}, \; \forall t \in \mathcal{T}, & \label{eq:gen_limits} \\
    & s_{g,t} - r_{g,t} = u_{g,t} - u_{g,t-1}, \; \forall t \in \mathcal{T}, & \label{eq:startup_shutdown} \\
    & \sum_{\tau=t}^{t+T_g^\text{up}-1} u_{g,\tau} \geq T_g^\text{up} \big( u_{g,t} - u_{g,t-1} \big), \notag \\
    & \forall t \in \{1, \ldots, T-T_g^\text{up}\}, & \label{eq:min_up_time} \\
    & \sum_{\tau=t}^{t+T_g^\text{down}-1} (1 - u_{g,\tau}) \geq T_g^\text{down} \big( u_{g,t-1} - u_{g,t} \big), \notag \\
    & \forall t \in \{1, \ldots, T-T_g^\text{down}\}, & \label{eq:min_down_time} \\
    & \theta_{i,t} - \theta_{k,t} = \frac{P_{ik,t}}{B_{ik}}, \; \forall (i,k) \in \mathcal{L}, \forall t \in \mathcal{T}, & \label{eq:dc_power_flow} \\
    & -P_{ik}^\text{max} \leq P_{ik,t} \leq P_{ik}^\text{max}, \; \forall (i,k) \in \mathcal{L}, \forall t \in \mathcal{T}, & \label{eq:line_limits}
\end{align}
\end{subequations}
where $u_{g,0} = 0$ for all $g$, $c_g(P_{g,t})$ is a cost associated with generator $g$ at time $t$ (for example, $c_g(P_{g,t}) = c_{2g} P_{g,t}^2 + c_{1g} P_{g,t} + c_{0g}$ for a polynomial cost function), $\mathcal{G}$ is a set of all generators, $\mathcal{L}$ is a set of all lines, and $\mathcal{T}$ is a set of time in hours.

The optimization problem in \eqref{eq:unit_commitment_main} is referred to as the UC problem, which minimizes the total operating cost over a given time horizon. The objective function in \eqref{eq:objective} comprises generation costs, startup costs, and shutdown costs.

The constraint~\eqref{eq:power_balance} ensures power balance at every time step, requiring total generation to match load demand, where \eqref{eq:gen_limits} enforces that the power output of each generator stays within its operational limits when the generator is online. The constraint~\eqref{eq:startup_shutdown} models the relationship between the on/off status of generators and their startup/shutdown states. In this setup, we consider ramping constraints as part of the startup and shutdown constraints, ensuring that generation changes smoothly during transitions. The constraints~\eqref{eq:min_up_time}--\eqref{eq:min_down_time} enforce minimum up-time and down-time requirements for generators, ensuring they remain online or offline for a specified duration after being switched on or off. The constraint~\eqref{eq:dc_power_flow} represents the linearized DC power flow equations, relating power flows to voltage angle differences and line susceptance. Finally, \eqref{eq:line_limits} ensures that power flow on each line does not exceed its thermal limits in either direction.

To extend this formulation and provide a baseline for comparison, the appendix presents a set of linear constraints based on the first-order Taylor approximation of voltage magnitudes and current flows.

\subsection{Unit Commitment using Conservative Bias Linear Approximations} \label{sub:uc_cbla}

The UC problem formulated in the previous subsection enforces network feasibility using a DC power flow approximation. While computationally efficient, the DC power flow model does not explicitly incorporate voltage and current constraints. In this section, we introduce penalty terms and replace the DC power flow constraints in~\eqref{eq:dc_power_flow}--\eqref{eq:line_limits} with a CBLA, which provides a linearized yet conservative representation of these constraints.

Let $\mathcal{B}$ be a set of all buses. The CBLA enforces voltage magnitude limits at each bus using the following inequalities:
\begin{subequations} \label{eq:voltage_CBLA_limit}
\begin{align}
    \underline{a}_{0,i,t} + \underline{\bm{a}}_{1,i,t}^T
    \begin{bmatrix}
        \bm{P_t} \\ \bm{Q_t}
    \end{bmatrix} \geq V_i^{\text{min}} - v_{i,t}^{\text{min}}, \; \forall i \in \mathcal{B}, \forall t \in \mathcal{T},
    \label{eq:voltage_lower} \\
    \overline{a}_{0,i,t} + \overline{\bm{a}}_{1,i,t}^T
    \begin{bmatrix}
        \bm{P_t} \\ \bm{Q_t}
    \end{bmatrix} \leq V_i^{\text{max}} + v_{i,t}^{\text{max}}, \; \forall i \in \mathcal{B}, \forall t \in \mathcal{T},
    \label{eq:voltage_upper}
\end{align}
\end{subequations}
where $v_{i,t}^{\text{min}}$ and $v_{i,t}^{\text{max}}$ are nonnegative penalty variables that allow violation of the voltage approximations at bus $i$ at time $t$. Here, $\bm{P}_t$ ($\bm{Q}_t$) denotes the vector of active (reactive) power injections at time $t$, while $V_i^{\text{min}}$ and $V_i^{\text{max}}$ are the lower and upper voltage magnitude limits at bus $i$. The coefficients $\underline{a}_{0,i}$ and $\underline{\bm{a}}_{1,i}$ correspond to the underestimating voltage approximation at bus $i$, and $\overline{a}_{0,i,t}$ and $\overline{\bm{a}}_{1,i,t}$ correspond to the overestimating voltage approximation.

Additionally, we enforce current flow limits in transmission lines through the following constraint:
\begin{equation}
    \overline{b}_{0,ik,t} + \overline{\bm{b}}_{1,ik,t}^T
    \begin{bmatrix}
        \bm{P_t} \\ \bm{Q_t}
    \end{bmatrix} \leq I_{ik}^{\text{max}} + z_{ik,t}, \; \forall (i,k) \in \mathcal{L}, \forall t \in \mathcal{T},
    \label{eq:current_CBLA_limit}
\end{equation}
where $z_{ik,t}$ is a nonnegative penalty variable allowing violation of the current flow approximation on line $(i,k)$ at time $t$, $I_{ik,t}^{\text{max}}$ denotes the maximum allowable current on line $(i,k)$, and $\overline{b}_{0,ik,t}$, $\overline{\bm{b}}_{1,ik,t}$ are coefficients for the overestimating current flow.

Let $F(\bm{P}_{g,t}, \bm{s}_{g,t}, \bm{r}_{g,t})$ denote the objective function in~\eqref{eq:objective}. With the inclusion of penalty terms, the modified objective becomes: 
\begin{equation}
    F(\bm{P}, \bm{s}, \bm{r}) + \sum_{t \in \mathcal{T}}\left(\lambda\sum_{(i,k) \in \mathcal{L}} z_{ik,t} + \gamma\sum_{i \in \mathcal{B}} (v_{i,t}^{\text{max}} + v_{i,t}^{\text{min}}) \right),
\end{equation}
where $\lambda$ and $\gamma$ are the penalty weights associated with violations of current and voltage magnitude constraints, respectively.

The CBLA method constructs linear constraints from coefficients derived at sampled operating points within a specified range. Before solving the UC problem, we can identify buses and lines that do not exhibit voltage or current violations in the sampled data; constraints for these elements remain inactive and can be excluded, reducing computational complexity without compromising feasibility.

By incorporating these constraints into the UC problem, CBLA ensures generation schedules remain feasible with respect to voltage and line flow limits while maintaining computational tractability. Unlike the standard DC power flow model, CBLA systematically accounts for these limits through conservative yet efficient linear approximations. Its sample-based nature also enables constraint screening, identifying likely binding constraints and allowing flexibility in selecting the error function depending on whether voltage or line flow limits are active.

To maintain the validity of CBLA-based constraints during the UC process, we adopt an iterative approach (Fig.~\ref{fig:algorithm}). In each iteration, the UC problem is solved using the current CBLA, and the resulting generation statuses and dispatch levels are checked against those from a first-order Taylor-based method or previous CBLA iterations. We also verify that power injections remain within the sampled region used to construct the CBLA. If violations are detected, new samples are drawn around the updated solution, the CBLA is recomputed, and the UC problem is resolved. This process mitigates extrapolation errors and prevents cycling between inconsistent UC solutions.

{\color{black} In addition, the search can be restricted to a limited set of candidate UC solutions generated by the solver. For instance, a small number of feasible candidates (e.g., up to 20) may be evaluated and compared based on feasibility and operational cost, with each candidate validated through an AC OPF evaluation. This approach provides a practical trade-off between solution quality and computational effort by selecting the best available candidate within the allowed computation time.} 

This CBLA-based formulation further allows two practical enhancements. First, it offers flexibility in controlling conservativeness by adjusting a bias term when constructing the constraints, enabling the user to trade off between feasibility margin and conservativeness. Second, it supports the inclusion of penalty terms that allow for controlled constraint violations, enhancing feasibility in cases where a strictly conservative formulation may be overly restrictive.

\subsection{Combinatorial Growth of Valid Generator Statuses}
\label{sec:valid_status_count}

In the UC formulation, each generator’s on/off status over time must satisfy minimum up and down time constraints~\eqref{eq:min_up_time}--\eqref{eq:min_down_time}, substantially reducing the number of admissible on/off sequences. Nevertheless, the feasible status space still grows exponentially with the number of time periods, making brute-force enumeration impractical.

To quantify this space, we implement an enumeration algorithm that counts binary vectors of length~$\abs{\mathcal{T}}$ satisfying a minimum run length~$d = T^\text{up} = T^\text{down}$ for both states, allowing exceptions for an initial short off-run and final short runs. Due to these exceptions, a clean closed-form expression is difficult to derive. Instead, the number of valid vectors is characterized recursively: sequences are built by appending valid runs (of at least length~$d$) to shorter valid sequences, naturally leading to a dynamic programming approach.

Let \( f(\abs{\mathcal{T}}, s, \ell) \) denote the number of valid binary sequences of length \( \abs{\mathcal{T}} \), ending in state \( s \in \{0,1\} \) with a current run of length \( \ell \). The recurrence relation is:
\[
f(\abs{\mathcal{T}}, s, \ell) = 
\begin{cases}
\sum\limits_{k=d}^{\floor{\sfrac{\abs{\mathcal{T}}}{2}}} f(\abs{\mathcal{T}} - k, 1 - s, k), \; \text{if } \ell = d, \abs{\mathcal{T}} \geq d \\[1ex]
f(\abs{\mathcal{T}} - 1, s, \ell - 1), \; \text{if } \ell > d \\[1ex]
1, \; \text{if } \abs{\mathcal{T}} = \ell \leq d \text{, one of the exceptions holds} \\[1ex]
0, \; \text{otherwise}
\end{cases}
\]
where the exceptions include:
\begin{itemize}
    \item Initial 0’s from position 1 with length less than $d$,
    \item Final 0’s or 1’s from position \( \abs{\mathcal{T}} - d + 2 \) onward.
\end{itemize}

The total number of valid sequences is
\[
F(\abs{\mathcal{T}}) = \sum_{s \in \{0,1\}} \sum_{\ell = 1}^{\abs{\mathcal{T}}} f(\abs{\mathcal{T}}, s, \ell).
\]

For example, with \( \abs{\mathcal{T}} = 5 \) and \( d = 2 \), applying the recurrence yields 13 valid sequences for a single generator. Therefore, for six generators, the total number of configurations is 
$(F(5))^6 = 13^6 = 4,\!826,\!809$,
demonstrating how the feasible status space grows rapidly even for modest systems. This growth illustrates why brute-force enumeration of all valid generator commitment combinations quickly becomes intractable. To manage this complexity, we avoid exhaustive search by solving the UC problem iteratively: each iteration only considers a manageable subset of decisions, and additional constraints are introduced in subsequent iterations to guide the solution away from unstable commitment patterns, such as cycling or oscillating generator decisions, i.e., repeatedly looping through statuses. The algorithm is explained in more detail in Section~\ref{sub:computation}.

\subsection{AC Optimal Power Flow with Load Shedding} \label{sub:AC_OPF_LS}

The AC Optimal Power Flow (AC OPF) problem determines the optimal generation dispatch by solving the full nonlinear AC power flow equations while satisfying network constraints. Using generator commitment statuses from the UC problem in~\eqref{eq:unit_commitment_main} (solved with formulations such as DC power flow, first-order Taylor approximations, or CBLAs), we compute the dispatch for each online generator at each hour. The multi-period formulation minimizes the total cost across all hours while ensuring power balance and network feasibility at each time. If constraints prevent meeting full demand, load shedding is used as a last resort, with $L_{i,t}$ denoting the load shed at bus $i$ and time $t$. The AC OPF with load shedding is formulated as follows:
\allowdisplaybreaks
\begin{subequations} \label{eq:acopf_ls}
\begin{align} 
    \min \ & \sum_{t \in \mathcal{T}} \left( \sum_{g \in \mathcal{G}} c_g(P_{g,t}) + \sum_{i \in \mathcal{B}} C^\text{shed}(L_{i,t}) \right) \label{eq:objective_acopf_ls} \\
    \text{s.t.} \  & [\forall i \in \mathcal{B}, \forall t \in \mathcal{T}, \forall g \in \mathcal{G}, \forall (i,k) \in \mathcal{L}] \nonumber \\
    & P_{i,t}^\text{gen} - P_{i,t}^\text{load} + L_{i,t} =  \nonumber \\
    & \quad \sum_{k \in \mathcal{B}} V_{i,t} V_{k,t} \big( G_{ik} \cos\theta_{ik,t} + B_{ik} \sin\theta_{ik,t} \big), \label{eq:real_power_balance_ls} \\
    & Q_{i,t}^\text{gen} - Q_{i,t}^\text{load} + \beta_i L_{i,t} = \nonumber \\
    & \quad \sum_{k \in \mathcal{B}} V_{i,t} V_{k,t} \big( G_{ik} \sin\theta_{ik,t} - B_{ik} \cos\theta_{ik,t} \big), \label{eq:reactive_power_balance_ls} \\
    & P_{g,t}^\text{min} u_{g,t} \leq P_{g,t} \leq P_{g,t}^\text{max} u_{g,t}, \label{eq:gen_limitsP_acopf_ls} \\ 
    & Q_{g,t}^\text{min} u_{g,t} \leq Q_{g,t} \leq Q_{g,t}^\text{max} u_{g,t}, \label{eq:gen_limitsQ_acopf_ls} \\
    & V_{i,t}^\text{min} \leq V_{i,t} \leq V_{i,t}^\text{max}, \label{eq:voltage_limits_ls} \\
    & S_{ik,t} = \sqrt{ P_{ik,t}^2 + Q_{ik,t}^2 } \leq S_{ik}^\text{max}, \label{eq:thermal_limits_ls} \\
    & 0 \leq L_{i,t} \leq P_{i,t}^\text{load}, \quad 0 \leq \beta_i L_{i,t} \leq Q_{i,t}^\text{load}, \label{eq:load_shed_limits_acopf}
\end{align}
\end{subequations}
where $\beta_i = \tan \alpha_i$, given that $\cos \alpha_i$ represents the power factor associated with load $i$.
The objective function in~\eqref{eq:objective_acopf_ls} minimizes the total cost of generation over all time periods while penalizing load shedding using a function $C^\text{shed}(L_{i,t})$ to discourage demand curtailment unless necessary. The power balance constraints in~\eqref{eq:real_power_balance_ls}--\eqref{eq:reactive_power_balance_ls} enforce the nonlinear AC power flow equations for active and reactive power at each bus and time period. The generator limits in~\eqref{eq:gen_limitsP_acopf_ls}--\eqref{eq:gen_limitsQ_acopf_ls} ensure that active and reactive power generation respects the UC status $u_{g,t}$. The voltage limits in~\eqref{eq:voltage_limits_ls} ensure that bus voltages remain within safe operating ranges. The thermal constraints in~\eqref{eq:thermal_limits_ls} restrict apparent power flows on transmission lines to their rated capacities. Finally, the load shedding constraints in~\eqref{eq:load_shed_limits_acopf} ensure that the curtailed demand remains non-negative and does not exceed the original load at any bus and time period. This formulation integrates the UC results into AC OPF, optimizing the dispatch of committed generators while allowing for load shedding with constant power factors.


\section{Numerical results} \label{sec:simulation}

In this section, we conduct numerical experiments on several test cases to examine the behavior of CBLA, highlight the benefits of error function design, and demonstrate the effectiveness of CBLA in a UC problem.

The test cases used in the simulations are the IEEE 24-bus system and the IEEE 30-bus system, both of which are accessible in M{\sc atpower}~\cite{zimmerman_matpower_2011}. For approximations of voltage and current flow, we draw $500$ samples by varying the power injections within a range of $70\%$ to $130\%$ of their nominal values. Both voltage and current flow values are reported in per unit (pu). We use the $\ell_1$ norm as the loss function $\mathcal{L}(\,\cdot\,)$. The numerical simulation was conducted in MATLAB using Yalmip~\cite{Lofberg2004} and in Julia using the Optim package. 

{\color{black} The number of samples (500–1000 in the reported experiments) is selected to provide adequate coverage of the multidimensional injection space while maintaining computational tractability. Empirically, we observe that beyond this range (e.g., more than 500–1000 samples for the tested systems), the marginal improvement in approximation accuracy and solution robustness becomes negligible relative to the additional computational cost.}

\subsection{Conservative Bias Linear Approximations} \label{sub:sim_CBLA}

We begin our numerical tests by examining the effects of changing the error function in~\eqref{eq:error_function} in our CBLA approach. As discussed in Section~\ref{sub:cbla_error_function}, error functions are designed to balance conservativeness and accuracy. By testing various error functions, we aim to understand their impact on the number of violated samples and the accuracy of the approximated power flow equations.

\begin{figure}[ht!] 
\vspace{-1.2em}
    \centering
    \subfloat[$g(\epsilon) = \epsilon^2, h(\epsilon) = \epsilon^2$.\label{fig:a1}]
    {\includegraphics[width=0.5\linewidth,height=0.44\linewidth]{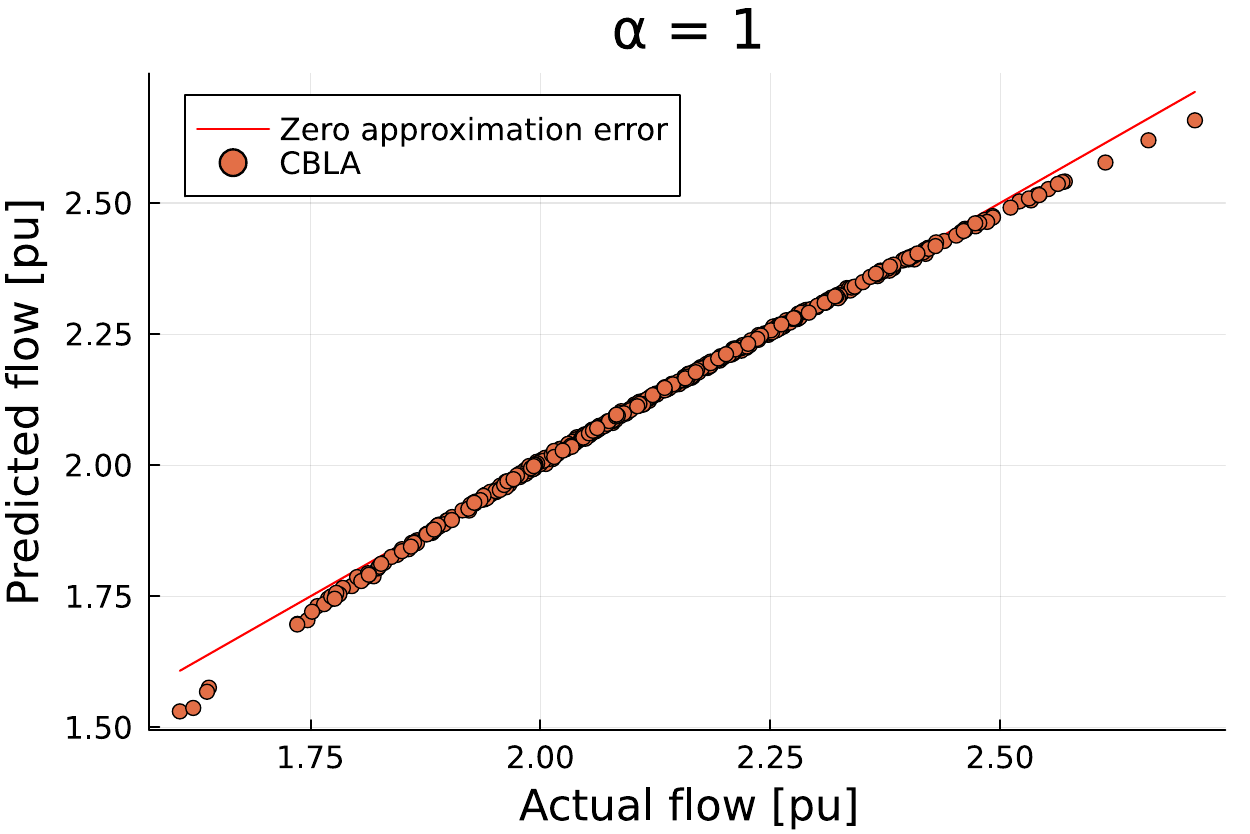}} \hfill
    \subfloat[$g(\epsilon) = \epsilon^2, h(\epsilon) = 10\epsilon^2$.\label{fig:a10}]
    {\includegraphics[width=0.5\linewidth,height=0.44\linewidth]{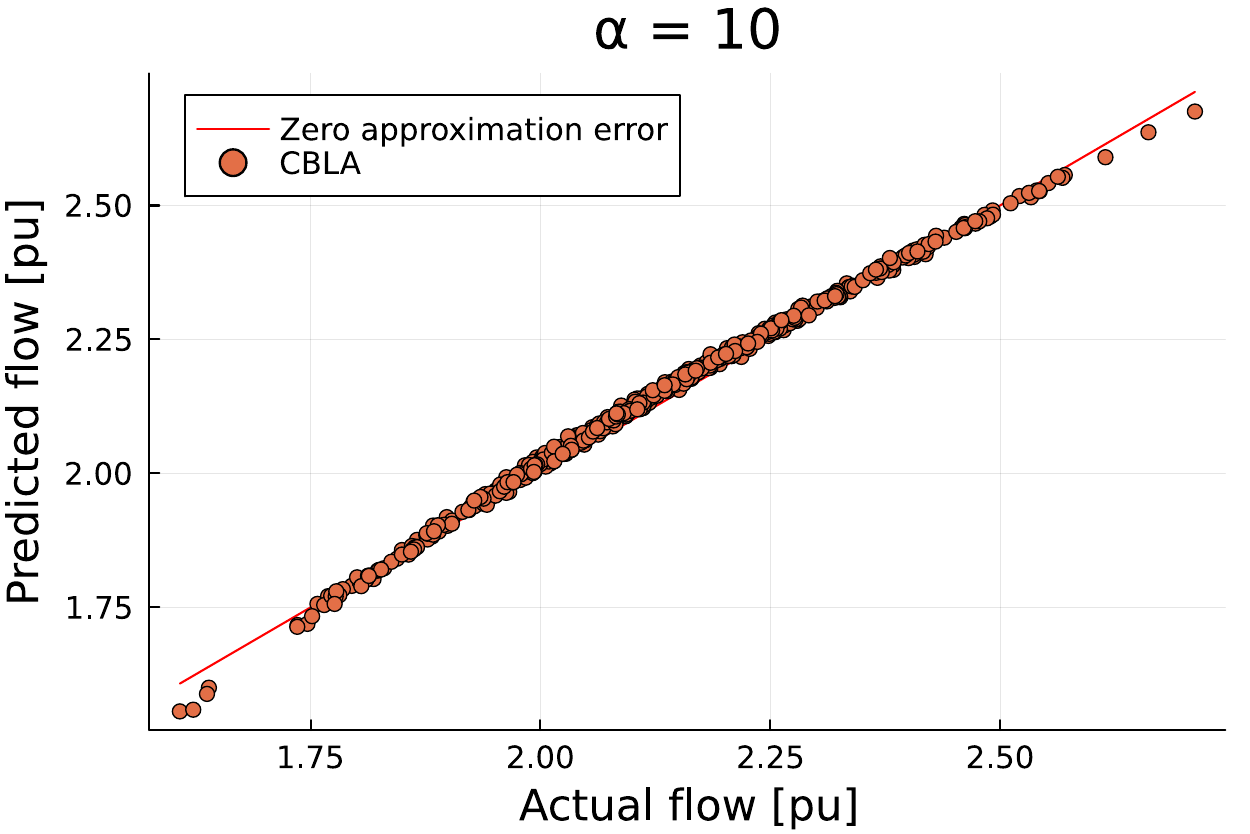}} \\[0.5em]
    
    \subfloat[$g(\epsilon) = \epsilon^2, h(\epsilon) = 100\epsilon^2$.\label{fig:a100}]
    {\includegraphics[width=0.5\linewidth,height=0.44\linewidth]{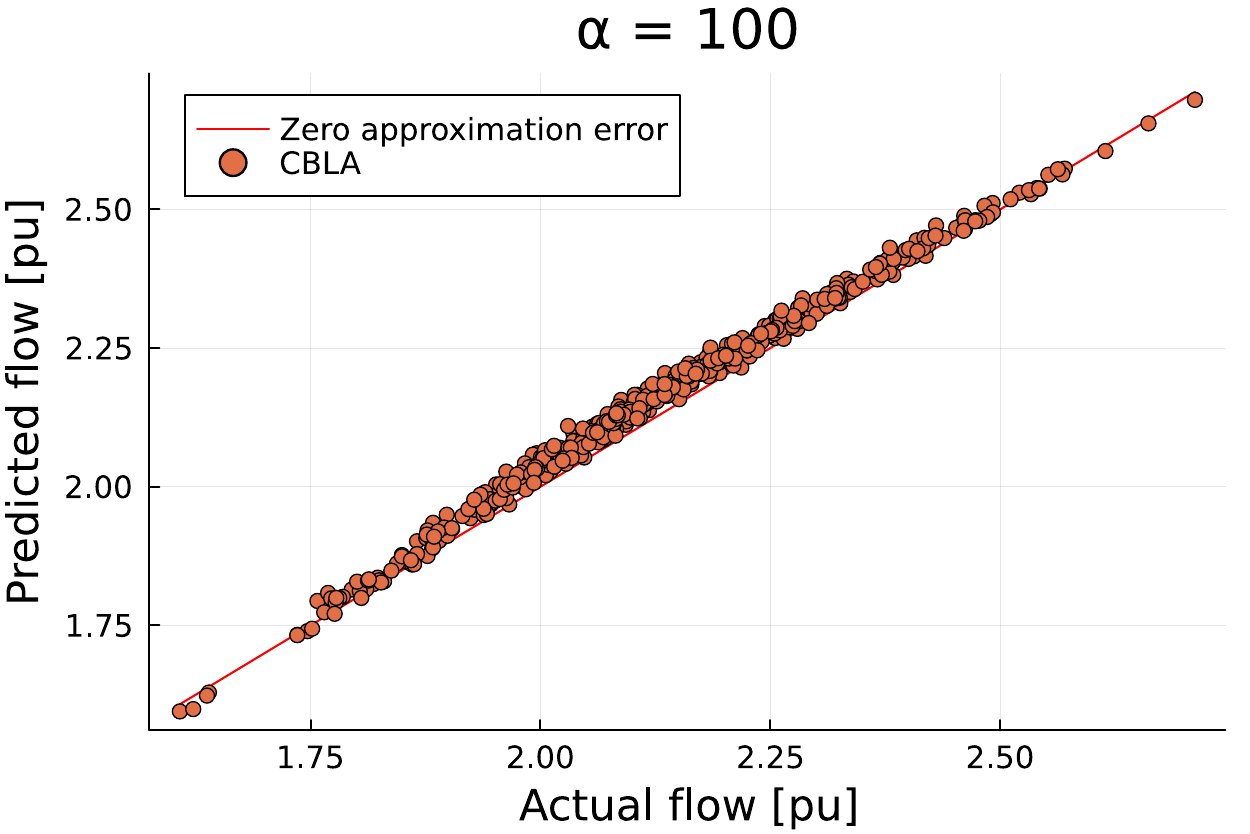}} \hfill
    \subfloat[$g(\epsilon) = \epsilon^2, h(\epsilon) = 1000\epsilon^2$.\label{fig:a1000}]
    {\includegraphics[width=0.5\linewidth,height=0.44\linewidth]{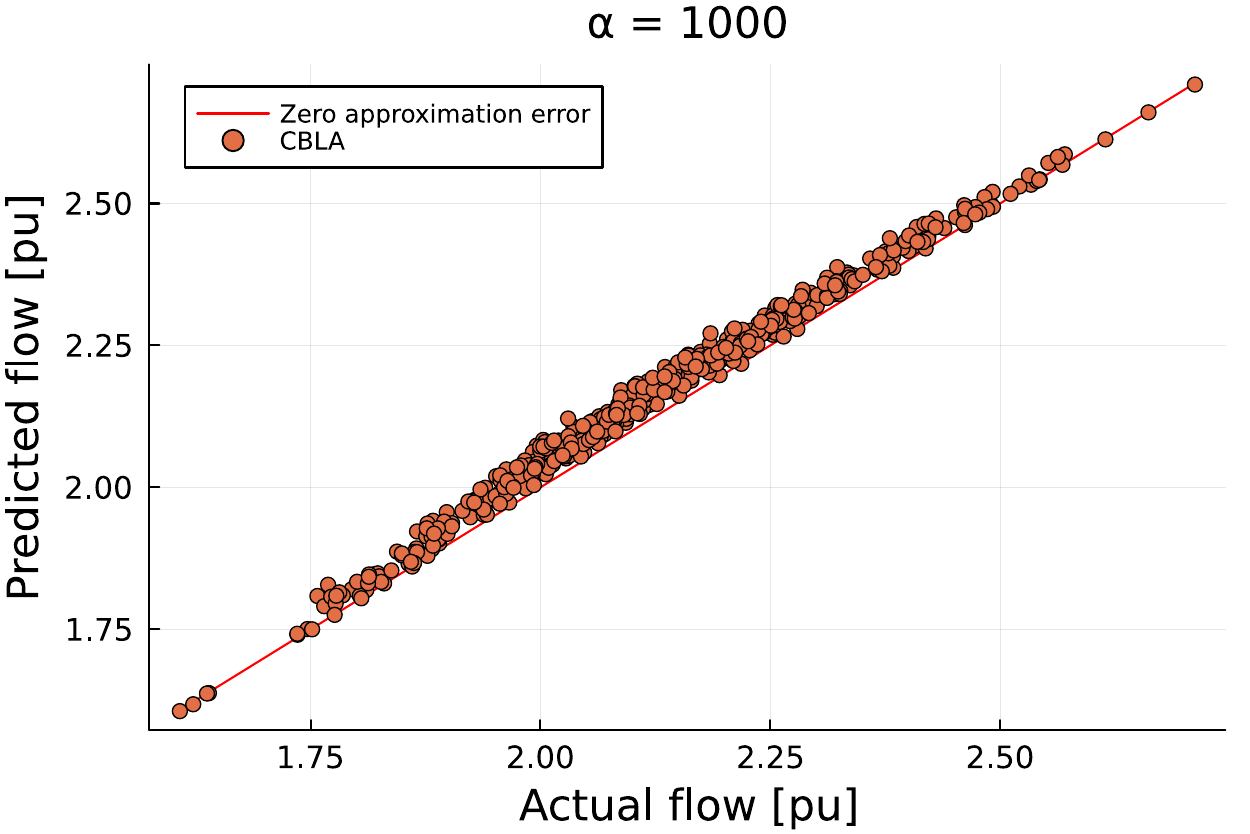}} 

    \caption{Plots of results when (a) $\alpha = 1$ (equivalent to the squared-$\ell_2$ loss), (b) $\alpha = 10$, (c) $\alpha = 100$, and (d) $\alpha = 1000$ for current flow from bus 3 to bus 24 in IEEE 24-bus system. The red points represent overestimating CBLAs. The black line represents a zero approximation error.}
    \label{fig:case24_CBLA} 
\vspace{-1.5em}
\end{figure}

In Fig.~\ref{fig:case24_CBLA}, we present the results of using the CBLA approach to intentionally overestimate the predicted current flow from bus 3 to bus 24 in IEEE 24-bus system using different quadratic error functions. The error functions used in this test are defined as $g(\epsilon) = \epsilon^2$ and $h(\epsilon) = \alpha \epsilon^2$, where $\alpha$ is a parameter that we vary across the test. Specifically, the parameter $\alpha$ modifies the bias in the CBLA formulation; a higher value of $\alpha$ increases the tendency toward conservativeness. We adjust $\alpha$ to take values of 1, 10, 100, and 1000. When $\alpha = 1$, the error functions are equivalent (i.e., $g(\epsilon) = h(\epsilon)$), implying no distinction in cost between violating and not violating conservativeness (i.e., the squared-$\ell_2$ loss). In this scenario, most samples are well approximated, but some fall below the zero approximation error line, indicating a deviation from the overestimating objective.

Increasing $\alpha$ reduces the number of samples falling below the zero approximation error line, indicating better adherence to the overestimation goal. However, this improvement sacrifices overall accuracy. At $\alpha = 1000$, where most samples are intentionally overestimated, the accuracy tends to be lower compared to other scenarios. This trade-off highlights the importance of carefully selecting the value of $\alpha$ to achieve a suitable balance between overestimation and accuracy.

\begin{figure}[ht!]
\vspace{-0.65em}
	\centering 
	\includegraphics[width=0.65\linewidth]{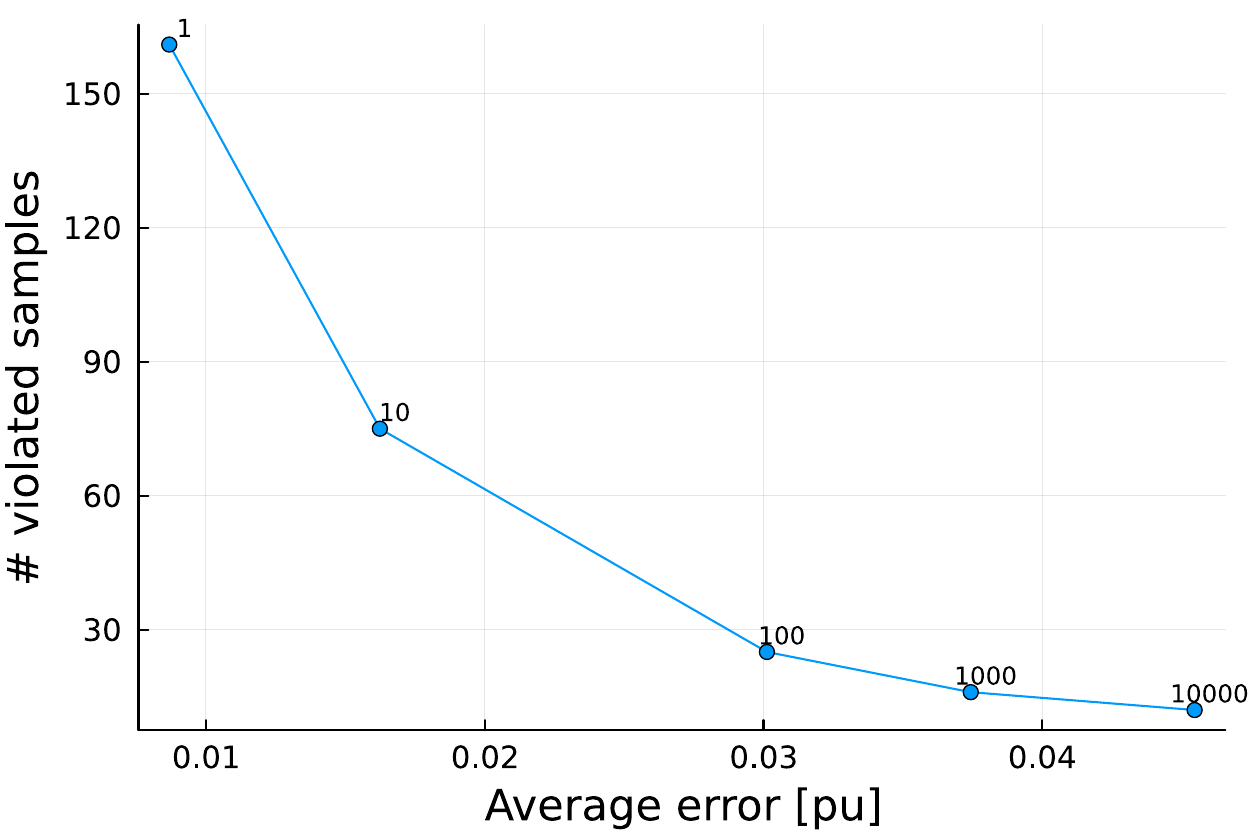} 
	\caption{Results showing the average error per sample in per unit (pu) and the number of violated samples due to overestimating CBLA of current flow from bus 3 to bus 24 in IEEE 24-bus system, as the value of $\alpha$ (labeled at each point) varies from $1$ to $10^4$.}
	\label{fig:case24_error_plot}
 \vspace{-1.1em}
\end{figure}

To gain further insight into the effects of varying $\alpha$, we plot the relationship between the average error per sample of the approximated flow and the number of violated samples when varying the value of $\alpha$ in Fig.~\ref{fig:case24_error_plot}. In this test, we adjust $\alpha$ over a range from $1$ to $10^4$. The results reveal a clear trend: as $\alpha$ increases, the average error per sample also increases while the number of violated samples decreases significantly, demonstrating the trade-off between conservativeness and accuracy. This is due to the increased enforcement of conservativeness in the error function. Specifically, the average error per sample increases from $0.00869$ when $\alpha = 1$ to $0.0455$ when $\alpha = 10^4$, while the number of violated samples decreases from $161$ when $\alpha = 1$ to just $12$ when $\alpha = 10^4$. 

\begin{table}[thb!] 
\caption{Approximated current flow errors and number of violated samples at representative buses in IEEE 24-bus system} \label{table:CBLA_error}
\begin{center}
\setlength\tabcolsep{3pt}
\begin{tabular}{c|c|c|c|c|c|c}
  Line & \multicolumn{3}{c|}{Average errors/sample} & \multicolumn{3}{c}{\# violated samples} \\
  \cline{2-7}
  (From-to)& $\alpha = 1$ & $\alpha = 10^2$ & $\alpha = 10^4$ & $\alpha = 1$ & $\alpha = 10^2$ & $\alpha = 10^4$\\
  \hline \hline
   3-14 & 0.00869 & 0.03012 & 0.04551 & 161  & 25  & 12  \\
  \hline
   6-10 & 0.00907 & 0.02274 & 0.03780 & 202 & 30 & 8  \\
  \hline
   9-12 & 0.01621 & 0.04961 & 0.09397 & 180 & 29 & 6   \\
  \hline
\end{tabular}
\end{center}
 \vspace{-1em}
\end{table}

The data in Table~\ref{table:CBLA_error} illustrates the relationship between the number of violated samples and the average approximated current flow errors across different $\alpha$ values at different lines. These results align with the trend observed in Fig.~\ref{fig:case24_error_plot}, confirming that as $\alpha$ increases, the average error per sample increases while the number of violated samples decreases.

\subsection{Application: Unit Commitment} \label{sub:sim_uc}

\begin{figure}[ht!]
\vspace{-0.6em}
	\centering 
	\includegraphics[trim={0cm 1.8cm 0cm 0.5cm},clip, width=0.76\linewidth]{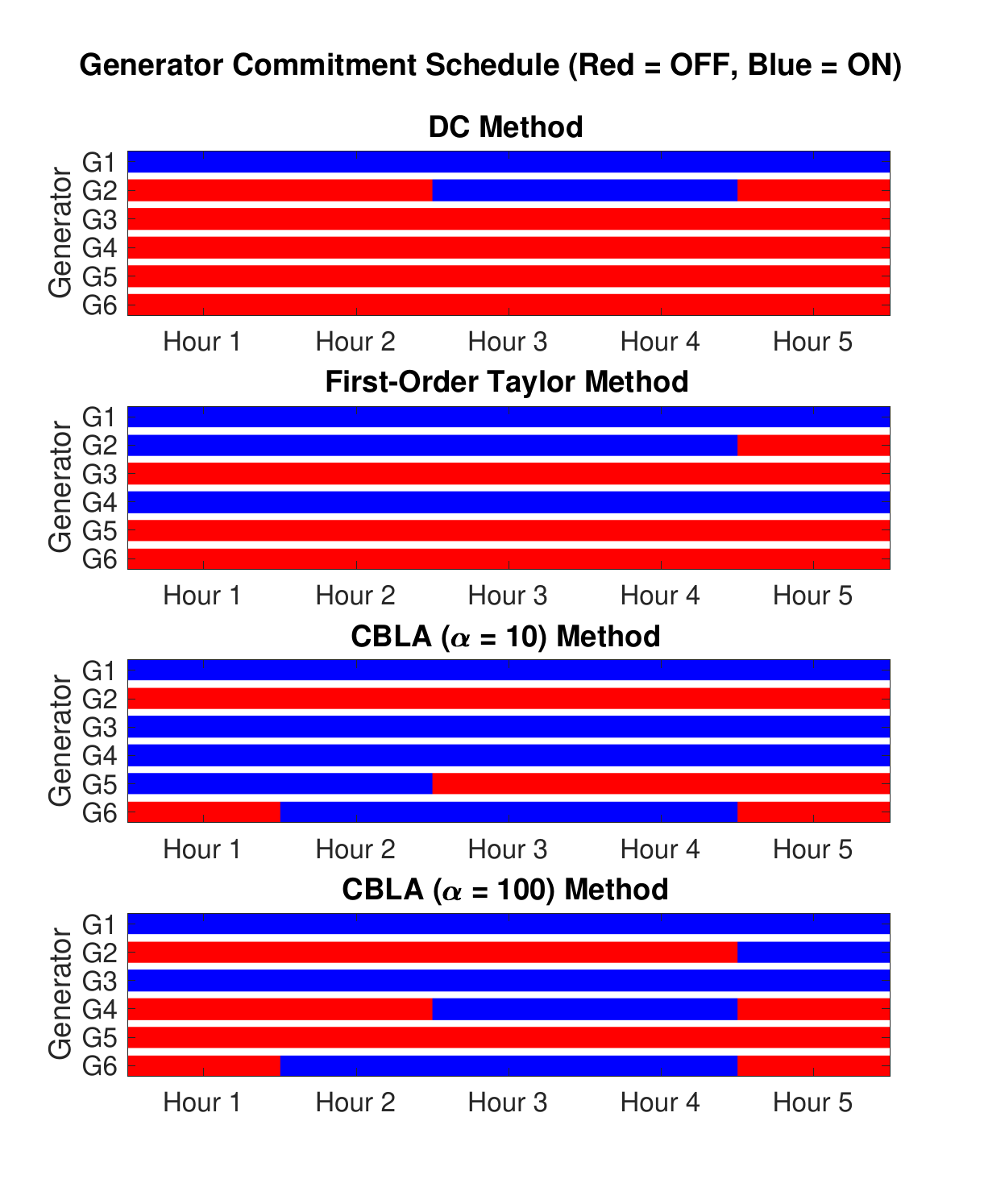} 
	\caption{Generator commitment schedules for the 30-bus system using three different power flow linearization methods: DC, first-order Taylor, and CBLA (when $\alpha = 10$ and $100$). Each row block corresponds to a generator across five scheduling hours. Blue indicates the generator is ON, while red indicates it is OFF.}
	\label{fig:bar_plot}
 \vspace{-0.7em}
\end{figure}

The comparison is conducted on the IEEE 30-bus test system, which has six generators. Voltage magnitude limits are set to 0.94~p.u. and 1.05~p.u. at all buses, and current magnitude limits are derived from the values in \texttt{mpc.branch(:,RATE\_A)} in the M{\sc atpower} case files. The cost of load shedding is $C^\text{shed}(L_{i,t}) = 500 L_{i,t}$. We draw $1000$ samples by varying power injections, both loads and generations, to cover possible load shedding and generation variation within a range of $70\%$ to $130\%$ of their nominal values. As shown in Table~\ref{tab:comparison_all_methods}, we report both the total cost, which includes load shedding penalties, and the generation-only cost, which excludes them to better isolate the impact of load shedding. Although the DC-based formulation often yields lower generation-only costs, it leads to much higher total costs due to frequent and severe load shedding. As illustrated in Fig.~\ref{fig:bar_plot}, the generator commitment schedules from CBLA differ from those of the DC and first-order Taylor methods, highlighting CBLA’s ability to adapt commitment decisions based on a more accurate power flow approximation.

This difference is especially evident in hour 4, where the DC-based approach leads to 11.82\% load shedding and a total cost of \$20{,}417. In comparison, the first-order Taylor approximation reduces load shedding to 5.01\%, with a much lower cost of \$11{,}683. CBLA reduces load shedding even further to just 1.92\%, while lowering the total cost to \$8{,}053.7---clearly demonstrating the value of its adaptive linearization.

\begin{table}[ht!]
    \centering
    \renewcommand{\arraystretch}{0.95}
    \setlength{\tabcolsep}{3pt}
    \small
    \caption{Hourly Comparison Across Different Approximations}
    \label{tab:comparison_all_methods}
    \begin{tabular}{l|l|c|c|c|c|c}

        & & \multicolumn{5}{c}{Hour} \\
        \cline{3-7}
        & & 1 & 2 & 3 & 4 & 5 \\
        \toprule

        \multicolumn{2}{l|}{\textbf{Load Demand (MW)}} 
        & 152.2 & 192.13 & 234.29 & 287.33 & 219.27 \\
        \midrule\midrule

        \multirow{9}{*}{\rotatebox[origin=c]{90}{\textbf{DC}}}
        & Load Shedding (\%) 
        & 0\% & 4.37\% & 2.57\% & 11.82\% & 15.23\% \\
        & \textbf{Outputs (MW)} & & & & & \\
        & \quad Gen 1 
        & 158.28 & 193.44 & 159.05 & 205.66 & 196.71 \\
        & \quad Gen 2 
        & 0 & 0 & 80 & 61.62 & 0 \\
        & \quad Gen 3 -- Gen 6 
        & 0 & 0 & 0 & 0 & 0 \\
        & Total Cost (\$)
        & \textbf{1817.6}
        & \textbf{6328} 
        & \textbf{6090} 
        & \textbf{20417} 
        & \textbf{18869} \\
        & Gen-Only Cost (\$)  
        & 1817.6 & 2135.2 & 3076.1 & 3431.5 & 2167.3 \\
        \midrule\midrule

        \multirow{9}{*}{\rotatebox[origin=c]{90}{\textbf{First-Order Taylor}}}
        & Load Shedding (\%) 
        & 0\% & 0\% & 0.70\% & 5.01\% & 2.42\% \\
        & \textbf{Outputs (MW)} & & & & & \\
        & \quad Gen 1 
        & 52.84 & 85.27 & 134.71 & 202.56 & 189.72 \\
        & \quad Gen 2 
        & 68.48 & 80 & 80 & 62.73 & 0 \\
        & \quad Gen 3 
        & 0 & 0 & 0 & 0 & 0 \\
        & \quad Gen 4 
        & 33.86 & 32.41 & 26.96 & 22.16 & 33.90 \\
        & \quad Gen 5 -- Gen 6 
        & 0 & 0 & 0 & 0 & 0 \\
        & Total Cost (\$)
        & \textbf{3483} 
        & \textbf{3682} 
        & \textbf{4800} 
        & \textbf{11683} 
        & \textbf{5875} \\
        & Gen-Only Cost (\$)
        & 3483.0 & 3682.1 & 3978.0 & 4480.5 & 3219.1 \\
        \midrule\midrule

        \multirow{11}{*}{\rotatebox[origin=c]{90}{\textbf{CBLA ($\alpha = 10$)}}}
        & Load Shedding (\%) 
        & 0\% & 0\% & 0\% & 1.92\% & 0\% \\
        & \textbf{Outputs (MW)} & & & & & \\
        & \quad Gen 1 
        & 58.14 & 58.32 & 99.31 & 152.81 & 127.40 \\
        & \quad Gen 2 
        & 0 & 0 & 0 & 0 & 0 \\
        & \quad Gen 3 
        & 26.96 & 27.47 & 45.73 & 46.77 & 50.00 \\
        & \quad Gen 4 
        & 50.40 & 55.00 & 53.42 & 50.02 & 47.48 \\
        & \quad Gen 5 
        & 18.73 & 24.61 & 0 & 0 & 0 \\
        & \quad Gen 6 
        & 0 & 29.04 & 40 & 40 & 0 \\
        & Total Cost (\$)
        & \textbf{4506.2} 
        & \textbf{5660.4} 
        & \textbf{4929.7} 
        & \textbf{8053.7} 
        & \textbf{3958.8} \\
        & Gen-Only Cost (\$)
        & 4506.2 & 5660.4 & 4929.7 & 5299.5 & 3958.8 \\
        \midrule\midrule

        \multirow{11}{*}{\rotatebox[origin=c]{90}{\textbf{CBLA ($\alpha = 100$)}}}
        & Load Shedding (\%) 
        & 0\% & 0\% & 0\% & 1.92\% & 0.93\% \\
        & \textbf{Outputs (MW)} & & & & & \\
        & \quad Gen 1 
        & 110.53 & 110.45 & 99.31 & 152.81 & 100.66 \\
        & \quad Gen 2 
        & 0 & 0 & 0 & 0 & 80.00 \\
        & \quad Gen 3 
        & 45.01 & 45.40 & 45.73 & 46.77 & 42.51 \\
        & \quad Gen 4 
        & 0 & 0 & 53.42 & 50.02 & 0 \\
        & \quad Gen 5 
        & 0 & 0 & 0 & 0 & 0 \\
        & \quad Gen 6 
        & 0 & 40.00 & 40.00 & 40.00 & 0 \\
        & Total Cost (\$)
        & \textbf{2637.0} 
        & \textbf{3799.1} 
        & \textbf{4929.7} 
        & \textbf{8053.7} 
        & \textbf{4830.8} \\
        & Gen-Only Cost (\$)
        & 2637.0 & 3799.1 & 4929.7 & 5299.5 & 3811.4 \\
        \bottomrule
    \end{tabular}
    \vspace{-0.5em}
\end{table}

While the DC and first-order Taylor methods rely on a single linearization of nonlinear power flows, CBLA builds a sample-based conservative approximation over a defined operating region, delivering significant improvements. Across all five hours, CBLA eliminates load shedding in four hours and limits it to minimal levels in the most challenging hour. CBLA consistently achieves the lowest total cost. For example, in hour 5, CBLA attains a lower total cost of \$3,958.8 compared to \$5,875 with the Taylor approximation and \$18,869 with the DC approximation.

The table also includes results for a more conservative CBLA variant with \( \alpha = 100 \). With \( \alpha = 100 \), the UC solution tends to avoid constraint violations. However, this added conservativeness comes at the cost of reduced accuracy in approximations. As a result, when the generation decisions from the UC problem are passed to the AC OPF, the resulting solution may still exhibit load shedding---particularly in Hour~5, where 0.93\% of the demand remains unmet.

{\color{black} This trade-off underscores the need to balance conservativeness and approximation accuracy. CBLA’s tunable parameter $\alpha$ allows for adjusting this balance to achieve the desired level of feasibility robustness and operational flexibility. By tailoring the approximation to system conditions, CBLA helps reduce constraint violations and yields UC solutions that generally remain feasible under the nonlinear AC power flow model.

A practical approach to selecting $\alpha$ is to evaluate constraint violations on an independent set of AC-feasible operating points and adjust the parameter accordingly. If $\alpha$ is chosen too small, the approximation may become insufficiently conservative, potentially leading to AC infeasibility in the UC solution. Conversely, excessively large values of $\alpha$ result in a more conservative formulation, which preserves feasibility but may increase operational cost. In either case, the impact primarily affects the optimality--conservativeness balance rather than the structural validity of the method.

Moreover, the benefits of CBLA become more pronounced in larger systems or under tight feasibility margins. Unlike the first-order Taylor approximation, which is inherently local, CBLA is constructed to remain conservative across a sampled region of interest. This range-aware property improves cost efficiency. Compared to traditional methods, CBLA provides a more adaptive solution framework that maintains the tractability of linear programming while reducing the gap between linear and nonlinear power system models.}

\section{Conclusion and Future Work} \label{sec:future work}

This paper presents a conservative bias linear approximation (CBLA) approach for approximating the power flow equations, aiming to balance conservativeness and accuracy while preserving linearity. The numerical results highlight the potential advantages of using CBLA for power flow problems. Selecting suitable error functions enables an effective balance between conservativeness and accuracy. Additionally, the ability to choose different error functions allows CBLA to be tailored to specific systems and operational conditions. When applied to the UC problem, the CBLA-based formulation consistently outperformed formulations based on DC power flow and first-order Taylor approximations, achieving lower overall operating costs in the subsequent evaluation of the AC optimal power flow. {\color{black} For benchmarking purposes, we compared CBLA against the DC power flow model and the first-order Taylor approximation, as these remain among the most widely adopted linearization approaches in both industry practice and academic research. In particular, DC power flow is the standard approximation for large-scale UC formulations, while Taylor-based linearizations are commonly used in optimization-based power system studies. A systematic comparison with other advanced linearized AC models (e.g., LPAC~\cite{Coffrin2014} and other formulations~\cite{fnt}) is a topic for future work.}

In addition, our future work aims to extend our current approach by developing additional CBLA through the use of piecewise linearizations. Moreover, we plan to apply our proposed approach to a broader range of power system planning and resilience tasks. This includes tackling complex bilevel problems, conducting capacity expansion planning studies, and extending the application of CBLA to unbalanced three-phase systems.

\section*{Acknowledgement}
\noindent P. Buason thanks the Department of Mechatronics Engineering at Rajamangala University of Technology Phra Nakhon.

\section*{Appendix: First-Order Taylor Approximation of Voltage and Current Constraints} \label{sec:Appendix}

Let $\bm{P}$ and $\bm{Q}$ denote the vectors of active and reactive power injections, respectively, and let $\bm{P}^{\text{nom}}$ and $\bm{Q}^{\text{nom}}$ represent their corresponding nominal values. The nominal operating point is determined independently for each time period according to the associated load conditions. Let $\mathcal{B}^{\text{PQ}}$ denote the set of PQ buses, and let $\widetilde{V}_{i,t}$ represent the linearized estimate of the voltage magnitude at bus $i$ during time period $t$. The superscript “nom” indicates that a quantity is evaluated at the nominal operating point.

For buses $i \in \mathcal{B}^{\text{PQ}}$, the voltage magnitude is approximated by a first-order expansion around the nominal operating point:
\begin{equation}
    \widetilde{V}_{i,t} = V_i^{\text{nom}} 
    + \left.\frac{\partial V_i}{\partial \bm{P}}\right|_{\text{nom}} \cdot (\bm{P}_t - \bm{P}^{\text{nom}})
    + \left.\frac{\partial V_i}{\partial \bm{Q}}\right|_{\text{nom}} \cdot (\bm{Q}_t - \bm{Q}^{\text{nom}}),
    \label{eq:volt_taylor}
\end{equation}
where the sensitivity matrices $\frac{\partial V_i}{\partial \bm{P}}$ and $\frac{\partial V_i}{\partial \bm{Q}}$ are evaluated at the nominal operating point using the inverse of the power flow Jacobian.

For buses outside the PQ set, the voltage magnitude is assumed to remain fixed at its nominal value:
\begin{equation}
    \widetilde{V}_{i,t} = V_i^{\text{nom}}, 
    \quad \forall i \notin \mathcal{B}^{\text{PQ}}.
    \label{eq:V_taylor}
\end{equation}

To approximate current limits, the real component of the apparent power flow on each line $(i,k) \in \mathcal{L}$ is expressed as a function of the difference between the linearized voltage magnitudes obtained from~\eqref{eq:V_taylor}:
\begin{equation}
    \Re(S_{ik,t}) \approx |Y_{ik}| \, (\widetilde{V}_{i,t} - \widetilde{V}_{k,t}),
\end{equation}
where $\Re(\cdot)$ denotes the real part of a complex quantity and $|Y_{ik}|$ is the magnitude of the line admittance.

Using this relationship, limits on the current magnitude can be approximated through the following linear constraints:
\begin{equation}
    -I_{ik}^{\text{max}} 
    \le |Y_{ik}| (\widetilde{V}_{i,t} - \widetilde{V}_{k,t}) 
    \le I_{ik}^{\text{max}},
    \quad \forall (i,k) \in \mathcal{L}, \; \forall t \in \mathcal{T}.
\end{equation}

Finally, to provide flexibility in the optimization model, slack variables can be incorporated to allow controlled violations of the linearized constraints, similar to the treatment used in~\eqref{eq:voltage_CBLA_limit}--\eqref{eq:current_CBLA_limit}.

\bibliographystyle{IEEEtran}
\bibliography{reference.bib}

\end{document}